\documentclass[twocolumn,prl,showpacs,preprintnumbers,amsmath,amssymb]{revtex4}
\usepackage{graphicx}
\usepackage{dcolumn}
\usepackage{bm}
\begin{document}

\title {A Unified Mechanism for Hydrogen Trapping at Metal Vacancies}

\author{Weiwei Xing$^1$}
\author{Xing-Qiu Chen$^1$}
\email[Corresponding author:]{xingqiu.chen@imr.ac.cn}
\author{Gang Lu$^2$}
\author{Dianzhong Li$^1$}
\author{Yiyi Li$^1$}

\affiliation{$^1$ Shenyang National Laboratory for Materials Science,
Institute of Metal Research, Chinese Academy of Sciences, Shenyang
110016, China}

\affiliation{$^2$ Department of Physics and Astronomy, California State University Northridge,
Northridge, California 91330, USA}

\date{\today}

\begin{abstract}

Interaction between hydrogen (H) and metals is central to many
materials problems of scientific and technological importance. Chief
among them is the development of H storage and H-resistant
materials. H segregation or trapping at lattice defects, including
vacancies, dislocations, grain boundaries, etc, plays a crucial role
in determining the properties of these materials. Here, through
first-principles simulations, we propose a unified mechanism
involving charge transfer induced strain destabilization to
understand H segregation behavior at vacancies. We discover that H
prefers to occupy interstitials with high pre-existing charge
densities and the availability of such interstitials sets the limit
on H trapping capacity at a vacancy. Once the maximum H capacity
is reached, the dominant charge donors switch from
the nearest-neighbor (NN) to the
next-nearest-neighbor (NNN) metal atoms.  Accompanying with this
long-range charge transfer, a significant reorganization energy would occur,
leading to instability of the H-vacancy complex.
The physical picture unveiled here appears universal across the BCC
series and is believed to be relevant to other metals/defects as well. The
insight gained from this study is expected to have important
implications for the design of H storage and H-resistant materials.

\end{abstract}


\maketitle

Interaction between H and lattice defects underlies diverse materials
phenomena \cite{book,Hmetal,review,Johnson}, including H storage
\cite{review,Chenp}, H embrittlement \cite{HE1,HE2,HE3}, metallic H
membranes \cite{Adhikari}, nuclear fusion reactors \cite{1,Zhou} and
H-assisted superabundant vacancy formation in metals
\cite{book,21,Fukai}, etc, to name a few. Crucial to all these
problems is trapping of H at the lattice defects, such as
vacancies, voids, dislocations, grain boundaries and cracks
\cite{vac1,20,Ohsawa,Ohsawa11,Luvoids,hd1,hd2,hd3,hd4}. In
particular, H trapping at vacancies has attracted the most attention
thanks to the facts that (1) vacancies are easier to study
than other defects but have profound influences on materials
properties; (2) vacancies hold many surprises and mysteries yet to
be explored; (3) the insight gained from vacancies can be applied to
other defects as well.

It is found that for BCC and FCC metals, up to six H atoms can be
trapped by a monovacancy in general because H prefers to occupy the
six octahedral ({\bf O}) interstitials surrounding the vacancy
\cite{book,Ohsawa,Ohsawa11}. However, there are exceptions - the
maximum H capacity can go up to 10 for Molybdenum (Mo)
\cite{25,Ohsawa,Ohsawa11} and 12 for Tungsten (W) and Aluminum
(Al)\cite{20}. Different interpretations of the available
experimental results have also been put forward: ($i$) The
competition between metal-H hybridization and the Coulombic
repulsion determines the position and number of H atoms at the
vacancy in BCC-Fe \cite{17}; ($ii$) Comparing to Fe, the greater H
trapping capacity in Al is attributed to a larger lattice constant
and more delocalized electronic states \cite{20}; ($iii$) In W, it
is found that the vacancy provides an isosurface of optimal charge
density that facilities the formation of H bubbles \cite{26}.
Clearly, an important question to ask is whether there exists a
unified mechanism of H trapping in metals?

\begin{figure}[htbp]
\begin{center}
\includegraphics[width=0.48\textwidth]{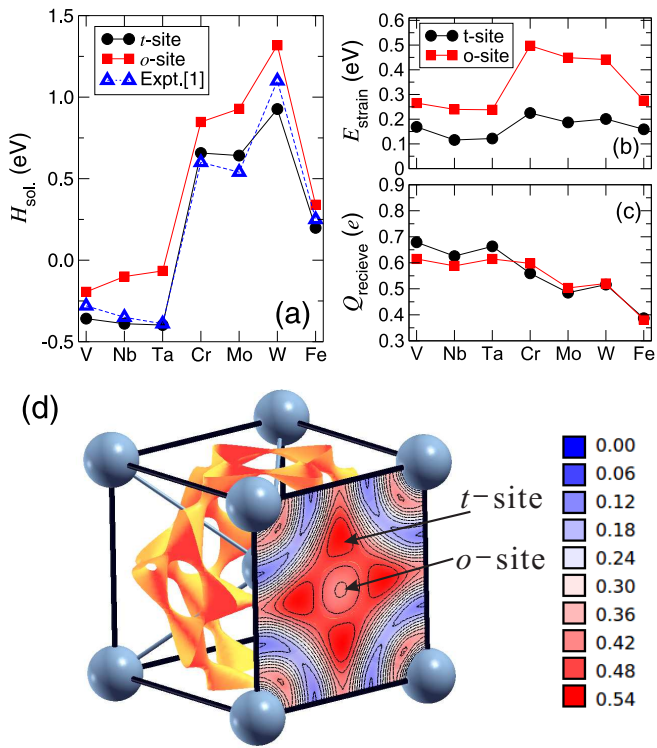}
\end{center}
\caption{ H in defect-free metals. (a) The solution enthalpies
(along with available experimental data \cite{book}) of H, (b) the
strain energy induced by H insertion and (c) the received charges of
H at {\bf T} or {\bf O}-sites in seven BCC-type metals (V, Nb, Ta,
Cr, Mo, W, and Fe). (c) Isosurface (with an isovalue of 0.42) of the
electron localized function (ELF) and its contour plot on (001)
plane showing the highest charge densities at the {\bf T}-sites in
V. The similar results have been observed for other six BCC metals
(Cr, Fe, Nb, Mo, Ta and W), but are not shown here. } \label{fig1}
\end{figure}

In this paper, we propose such a unified mechanism based on
quantum-mechanical density functional theory ({\small DFT})
calculations. We show that charge transfer induced strain
destabilization is responsible for H trapping behavior at vacancies
across the BCC series. By examining H trapping at monovacancies in
these BCC metals (including V, Nb and Ta as good candidates for H
storage; Mo and W as the most promising plasma facing materials of
nuclear fusion reactors, and Fe and Cr as the most commonly used
elements in H-resistant structural steels), we discover that H
prefers to occupy the interstitials with high pre-existing charge
densities with or without a vacancy. The availability of such
interstitials determines the H capacity at a vacancy. When the
maximum H capacity is reached, the dominant charge donors switch
from the NN to the NNN metal atoms. The long-range charge transfer
leads to a significant strain energy cost. The correlation between
the charge transfer and the onset of mechanical ``instability"
appears universal across the BCC series, and ultimately determines
the limit on H storage in metals.

The DFT calculations were carried out using Vienna \emph{ab}\emph{
initio} Simulation Package(VASP)\cite{Kresse1,Kresse2,Kresse3} with
the projector augmented wave potential(PAW) method
\cite{Blochl,Kresse4} and the generalized gradient approximation
(GGA) of Perdew-Burke-Ernzerhof(PBE)\cite{Predew} form was employed
for electron exchange and correlation. An energy cutoff of 400eV was
chosen for the plane-wave expansion of eigenfunctions. All results
were obtained with a 54-atom supercell; for Ta, Mo and W, a 128-atom
supercell was also used to check the convergence of the results. The
difference in H trapping energies between the 128-atom and 54-atom
supercells is less than 0.03 eV, confirming our results are
converged. The Brillouin-zone integrations were performed by using a
4$\times$4$\times$4 k-mesh for 54 atoms and 3$\times$3$\times$3
k-mesh for 128 atoms according to the Monkhorst-Pack
scheme.\cite{Monkhorst} Spin polarized calculations were performed
for Cr and Fe although the magnetism only slightly affects the
properties of Cr-H system.\cite{Ohsawa} We did not consider
zero-point energy (ZPE) correction in the calculations for the same
reason as explained in Ref. \cite{Ohsawa,Ohsawa11,17,26}.

To determine the strain energy of the vac-\emph{n}H complexes, we
remove H atoms and compute the total energies of H-free vacancy
with and without relaxing the metal atoms. The strain energy is
defined as the total energy difference before and after the
relaxation:
\begin{math}
E_{\textrm{strain}}(\textrm{vac-\emph{n}H}) =
E_{\textrm{hollow}}^{\textrm{unrelaxed}}-E_{\textrm{hollow}}^{\textrm{relaxed}}
\end{math}.
The differential strain energies is defined as the strain energy
difference between vac-(\emph{n}+1)H cluster and vac-\emph{n}H
cluster:
\begin{math}
\Delta E_\textrm{strain} =
E_{\textrm{strain}}(\textrm{vac-(\emph{n}+1)H})-E_\textrm{strain}(\textrm{van-\emph{n}H})
\end{math}.
The differential strain energy represents the change in the strain
energy induced by an additional H atom incorporated to the
vac-\emph{n}H cluster.

The charge deficit on the nearest-neighbor (NN) and the
next-nearest-neighbor (NNN) metal atoms surrounding the vacancy can
be estimated by the following equation:
\begin{math}
Q_{\textrm{losing}}^{X} =
\Sigma{[Q_{\textrm{Vac-}n\textrm{H}}^{X}(\textrm{Tm})-Q_{\textrm{Vac}}^{X}(\textrm{Tm})]}
\end{math},
where \emph{X} denotes the NN or NNN atoms surrounding the
monovacancy, and $Q_{Vac-nH}$ and $Q_{Vac}$ refer to the charges of
the NN or NNN atoms after and before the trapping of H,
respectively.


{\em H trapping in a perfect lattice} In defect-free BCC metals
considered here, we reproduce the known fact that H prefers the
tetrahedral interstitial ({\bf T}) to the octahedral interstitial
({\bf O}), as evidenced in Fig. \ref{fig1}a. The solution enthalpies
of H at the {\bf T}-sites are in excellent agreement with the
experimentally measured data in the low concentration limit of H
\cite{book}, verifying the reliability of our DFT calculations. To
our surprise, the strain energy induced by H occupying the {\bf O}
interstitial is much higher than the {\bf T} interstitial
(\emph{c.f.}, Fig. \ref{fig1}b). Being smaller than the interstitial
volume, H is not expected to yield substantial strain energy
difference between the {\bf T} and {\bf O} sites \cite{Zhou}. Thus
the result suggests that there is more to H trapping than a simple
consideration of size effect. Indeed, an inspection of the charge
density plot in Fig. \ref{fig1}d reveals a surprising clue. The
contours of the electronic localization function (ELF) \cite{ELF}
indicate that the {\bf T} interstitials exhibit the highest charge
density with an ELF of 0.54, much larger than that at the {\bf O}
interstitials with an ELF of 0.28. More interestingly, this
observation appears universal across BCC series examined here.
Utilizing the Bader's technique \cite{Bader}, we find that H atoms
trapped at both interstitials attract electrons from the host atoms
in nearly the same amount (see Fig. \ref{fig1}c). Because the
pre-existing charge density at the {\bf O} interstitials is much
lower than that at the {\bf T} interstitials, transferring the same
amount of charge to the H atoms at the {\bf O} sites would lead to a
higher strain energy, thereby rendering the {\bf O}-sites less
favorable. The fact that the most favorable H trapping sites
coincide with the highest pre-existing charge density across the BCC
series suggest that the charge accumulation plays a crucial role in
H trapping.

\begin{figure*}[htbp]
\centering
\includegraphics[width=0.80\textwidth]{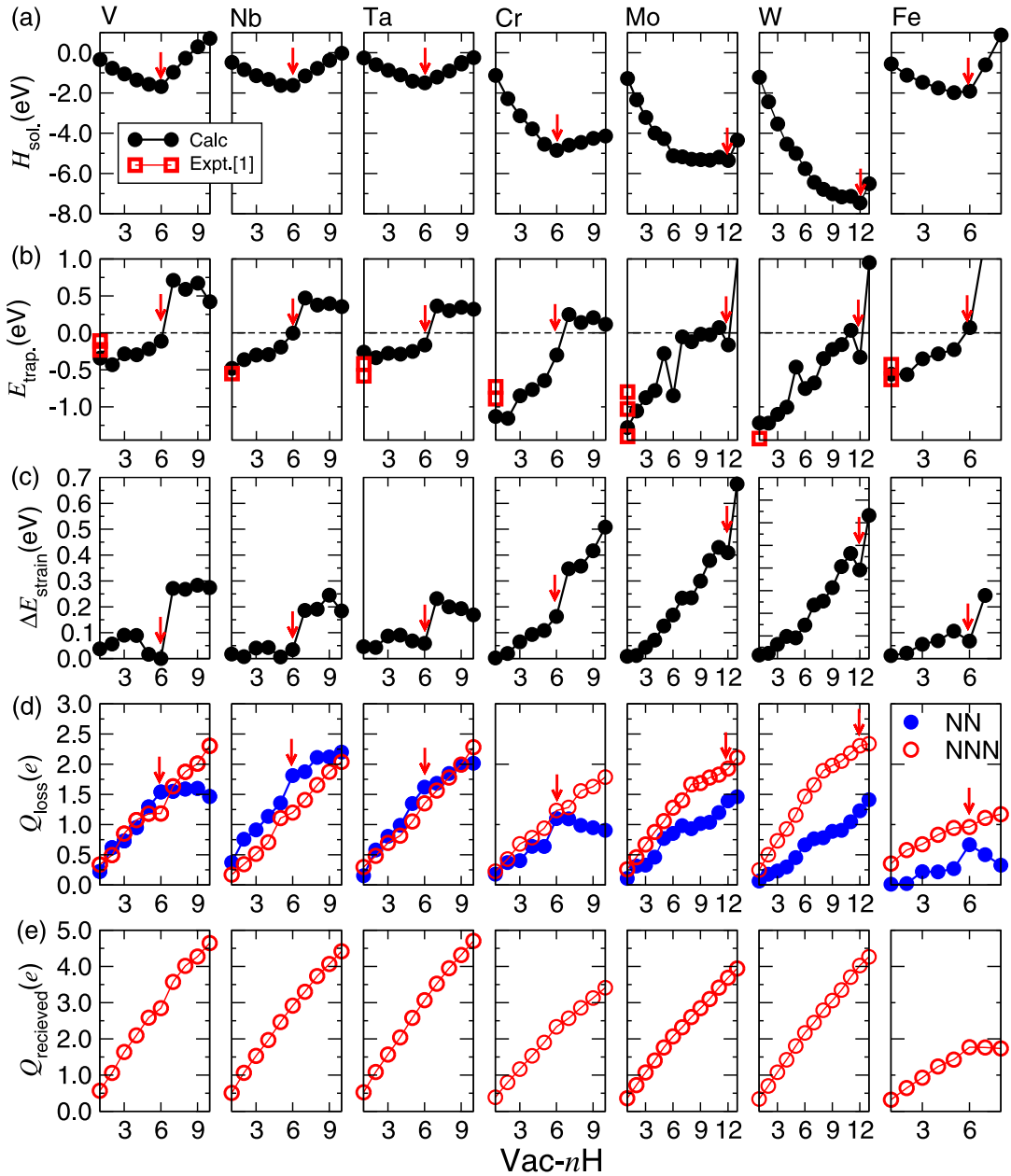}
\caption{ H trapping at a monvacancy across the BCC series. (a)
Solution enthalpies $H_{sol}$, (b) H trapping energies (along with
available experimental data), (c) the differential strain energies,
(d) the charge transfer from the eight NN (solid circles) and the
six NNN (hollow circles) metal atoms, and (e) the received charges
on H as a function of the number of H. Note that, in panel (b) the
zero enthalpy corresponds to the energy of an interstitial H at a
tetrahedral site as far as possible away from the vacancy. In panel
(b) we also list the experimental trapping energies  for V (-0.11 eV
and -0.23 eV), Fe (-0.63 eV and -0.43 eV), Ta (-0.42 eV), Mo (-1.03
eV and -0.80 eV) and W (-1.43 eV) measured by the ion
implantation/channeling method, Cr (-0.89 eV and -0.73 eV) measured
by the superabundant vacancy formation/thermal desorption
spectroscopy, and Nb ( -0.55 eV), Mo (-1.4 eV) and Ta (-0.58 eV)
measured by the positron annihilation spectroscopy
\cite{book,Hmetal}. Although the exact number of H at the vacancies
was uncertain in the experiments \cite{book,Hmetal}, the measured
data are generally in a good agreement with the simulation results.
\label{fig2}}
\end{figure*}

{\em H trapping at vacancies} When a vacancy, the most common
defect, is introduced in metals, the trapping behavior of H is quite
different from that in defect-free metals
\cite{book,Ohsawa,Ohsawa11,17,20,26}. Here, we denote a H-vacancy
complex as vac-$n$H in which \emph{n} denotes the number of
segregated H atoms at the monovacancy. As illustrated in Fig.
\ref{fig2}(a and b), we reproduce well the previous findings
\cite{Ohsawa,Ohsawa11} for seven BCC metals. A maximum of six H
atoms can be trapped at a monovacancy in V, Nb, Ta, Cr and Fe (group
I), whereas up to 12 H atoms can be accommodated at a monovacancy in
Mo and W (group II).

\begin{figure*}[htbp]
\begin{center}
\includegraphics[width=0.98\textwidth]{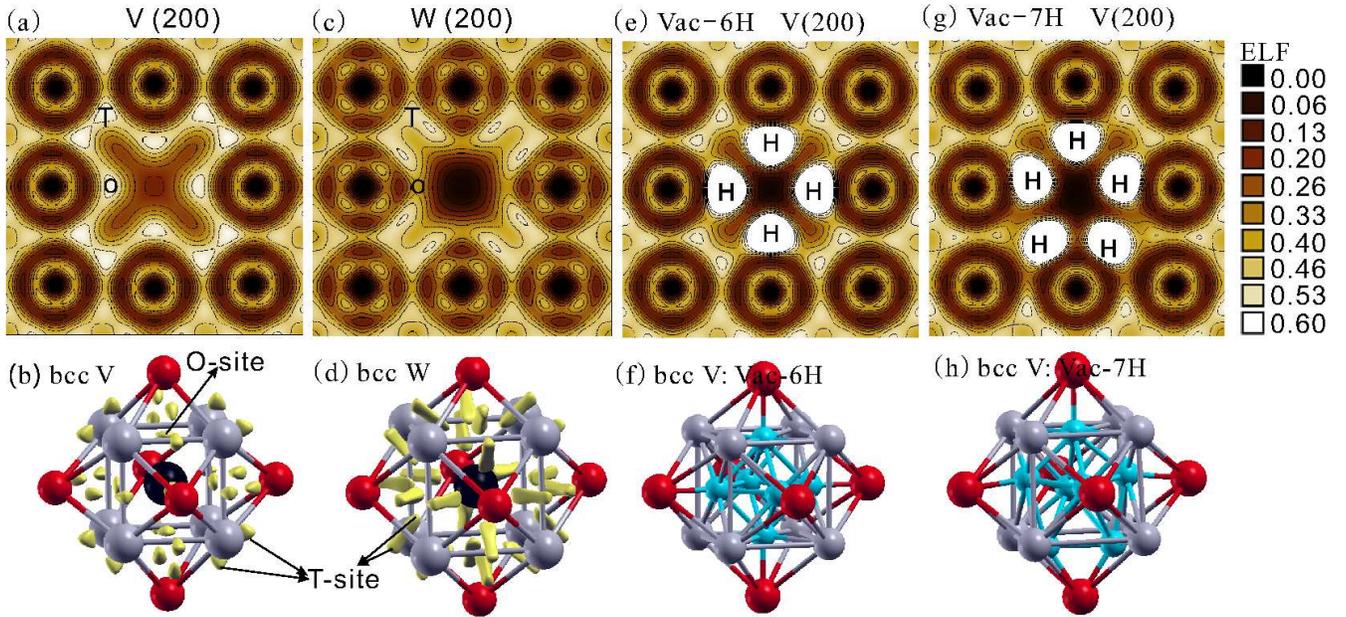}
\end{center}
\caption{The change of charge density at the interstitials before
and after H is introduced in V and W. The contours of the electron
localization functions (ELFs) for the (200) plane and the atomic
structure surrounding the vacancy. Panels (a,b): pure V; Panels
(c,d): pure W; Panels (e,g): six H atoms are trapped at the vacancy
(vac-6H) in V; Panels (f,h): seven H atoms at the vacancy leading to
instability in V. The 3D ELF isosurfaces in V (with an isovalue of
0.55) in Panel (b) and W (with an isovalue of 0.41) in Panel (c)
evidence the vacancy-induced highest charge accumulations at the
interstitial \emph{o}- or \emph{t}-sites for the H-free cases. The
large black, grey, red and green spheres denote the vacancy, the NN,
NNN metal atoms and the trapped H atoms, respectively. \label{fig3}}
\end{figure*}

{\em Charge transfer induced strain destabilization} To elucidate H
trapping behavior at vacancies, we choose V with more localized
3\emph{d} electronic states and W with less localized 5\emph{d}
states as examples to analyze the charge distribution. In Fig.
\ref{fig3}, we display the ELF contours on (200) plane for V and W.
For V, we find that the highest charge density occurs at {\bf O} and
{\bf T}-sites, similar to the defect-free case. In particular, the
six {\bf O} sites surrounding the vacancy now have approximately the
same (high) charge density as the T sites. The same result is also
found for other group I metals (Nb, Ta, Cr and Fe). In contrast, for
group II metals (Mo and W), the highest charge density takes place
at twelve equivalent tube-like regions connecting the NN {\bf
T}-sites as shown in Fig. \ref{fig3}(c,d)).

Similar to the defect-free case, H is also found to occupy the
interstitials with the highest charge densities. For V, up to six H
atoms can be accommodated at a vacancy and they occupy the six {\bf
O}-sites with the highest pre-existing charge density as shown in
Fig. \ref{fig3}(e,f). When the seventh H atom is introduced, all
atoms have to reorganize themselves; this reorganization proceeds in
such a way that additional interstitials with high charge densities,
such as the nearby {\bf T}-sites, become available to H occupation.
Since the {\bf T}-sites are inequivalent from the {\bf O}-sites,
occupation of these {\bf T} sites would introduce distortion to the
lattice. Indeed as shown in Fig. \ref{fig2}c, there is a jump of the
strain energy at $n=7$. In fact, this sudden increase in the strain
energy is observed for all group I metals. Moreover, this charge
transfer induced strain destabilization is also found for Mo and W
of the group II. As shown in Fig. \ref{fig2}(a-d), the similar jump
in the strain energy is evident for $n=13$. In this case, H
occupation at additional interstitials beyond the twelve
cylinder-like sites leads to lattice distortion and increases the
strain energy. This observation underlies the fact that up to twelve
H atoms can be trapped at a vacancy in the group II metals.

As displayed in Fig. \ref{fig2}(e), the total charges that the
vac-$n$H complexes received are monotonously increased as a function
of $n$. Obviously, the transferred charges have to mainly originate
from the metallic atoms surrounding the vacancy, including eight NN
atoms and six NNN atoms. Whereas the transferred charges from the
NNN atoms increase monotonously as $n$ across the BCC series, the
charge transfer from the NN atoms initially increase, but after a
kink (as marked by arrows in Fig. \ref{fig2}(d)) at \emph{n} =6, it
starts to decrease for the group I metals, V, Cr and Fe. Although
not as obvious as in V, Cr and Fe, the similar trend can be also
found at \emph{n} =6 for Ta and Nb in the group I. Interestingly,
the kink appears at \emph{n} = 6, corresponding to the maximal H
capacity for the group I metals. Hence the declining charge transfer
from the NN atoms gives rise to the instability of the vacancy-$n$H
complex. In contrast, such kinks are not observed for the group II
metals.

For the type-I group, the charge transfer to H is relatively more
difficult due to more localized nature of the electron states. In
this case, the charge transfer from the NNN atoms becomes
increasingly important as the H-vacancy cluster grows. For some
instances, with the increasing charge transfer from the NNN atoms,
the charge contribution from the NN atoms starts to decrease,
leading to the kink as observed at \emph{n}=6. However, this
long-range charge transfer from the NNN atoms yields a higher
reorganization energy. Therefore, the kink in the charge transfer
also coincides with the sharp increase of the strain energy and the
onset of the instability. For the group II, the charge transfer not
only stems from the NN atoms but also the NNN atoms thanks to the
more delocalized nature of electronic states. Thus there is an
obvious transition from the NN atoms to the NNN atoms as in the
group I, and no apparent kinks in the charge transfer for the group
II.

Finally, we note that H also prefers to occupy the interstitials
with high pre-existing charge density in other types of lattice
defects, such as grain boundaries \cite{XZ2010}, dislocations
\cite{XZ2010,Lu2006}, and cracks \cite{YS2013}. We have also
revisited the case of FCC Al and the similar charge transfer induced
strain destabilization is also observed. This universal
behavior of H occupation at defects accentuates the importance of
the proposed mechanism in understanding H segregation in metals.

In summary, we find that H prefers to occupy interstitials with high
pre-existing charge densities. The charge transfer induced strain
destabilization is responsible for the maximum H storage in BCC and
possibly other metals. The insights gained from the study could
provide guidance in the design of H storage or H-resistant
materials. Strain engineering, defect engineering and alloying could
be effective means to modify the charge density distribution at the
defects, thereby changing the H concentration in the material.

\textbf{Acknowledgements} This work was supported by the ``Hundred
Talents Project'' of the Chinese Academy of Sciences and from the
Major Research Plan (Grand Number: 91226204) of the National Natural
Science Foundation of China (Grand Numbers: 51074151 and 51174188)
and Beijing Supercomputing Center of CAS (including its Shenyang
branch) and Vienna Scientific Clusters.

\end{document}